\documentstyle[12pt]{article}
\def\Journal#1#2#3#4{{#1} {\bf #2}, #3 (#4)}

\def\NCA{{\em Nuovo Cimento} A}

\def\NPB{{\em Nucl. Phys.} B}
\def\NP{{\em Nucl. Phys.}}
\def\NPPC{{\em Nucl. Phys. Proc. Suppl.} B}
\def\PLB{{\em Phys. Lett.}  B}
\def\PRL{\em Phys. Rev. Lett.}
\def\PRD{{\em Phys. Rev.} D}
\def\ZPC{{\em Z. Phys.} C}
\def\JETPL{{\em JETP Lett.}}
\def\YaF{{\em Yad. Fiz.}}
\def\SJNP{{\em Sov. J. Nucl. Phys.}}
\def\JMP{{\em J. Math. Phys.}}
\def\PR{{\em Phys. Rev.}}
\def\PZh{{\em Pisma Zh. Eksp. Teor. Fiz.}}
\def\ZhETF{{\em Zh. Eksp. Teor. Fiz.}}
\def\JETP{{\em Sov. Phys. JETP}}
\def\AP{{\em Ann. of Phys.}}
\def\HPA{{\em Helv. Phys. Acta}}
\def\IJMPA{{\em Int. J. Mod. Phys.} A}
\def\MPLA{{\em Mod. Phys. Lett.} A}

\def\be{\begin{equation}}
\def\ee{\end{equation}}
\def\bea{\begin{eqnarray}}
\def\eea{\end{eqnarray}}

\begin{document}
\begin{flushright}
Preprint INR-0926/96\\
July 1996
\end{flushright}
\begin{center}
{\bf The generalized Crewther relation: the peculiar aspects\\
of the analytical perturbative QCD calculations.}
\end{center}
\begin{center}            
{ A.~L~.Kataev }
\end{center}
\begin{center}
{\em Institute for Nuclear Research of the Academy of Sciences
of Russia,\\ 117312 Moscow, Russia}
\end{center}
\begin{center}
\vskip .5cm
{\bf Abstract}\\
{We summarize the current status of our
understanding of the structure of the
perturbative QCD expressions for the
QCD generalizations of the Crewther relation.}
\end{center}
\vskip 5 cm
Invited talk at the Workshop ``Contenuous Advances in QCD'',
Minneapolis, March 28-31, 1996

\newpage
\section{Introduction}

From time to time the detailed study of the results of the
{\bf analytical} multiloop calculations allow one to reveal
the existence of the internal
symmetries and of the definite properties
of the gauge theories under investigation.
In fact these typical features of the concrete models can be hidden
in the explicit
expressions of the coefficients of the perturbative series for the
renormalization group quantities. The classical example of the
immediate influence of the outcomes of the analytical calculations to
the further development of the understanding of the structure of the
perturbative series as the whole is
provided by the evaluation of the 3-loop contribution to the
$\beta$-function of the $N=4$ supersymmetric Yang-Mills
theory~\cite{AVT}. Indeed, in Ref.[1] it was found by the direct
diagram-by-diagram calculation that the corresponding $\beta$-function
is zero at the 3-loop order. In its turn, this foundation pushed ahead
theoretical works, which resulted in the formulation of the proof
of the validity of this  interesting property of the $N=4$ supersymmetric
gauge model
in all orders of the perturbation theory (see e.g.
Ref.[2]). Therefore, in spite of the fact that the calculation of
the related $\beta$-function coefficients presumes the introduction of
the regularization (definitely speaking, supersymmetric one) and thus
the renormalization procedure, the property of the conformal symmetry
is preserved in this model.

However, it is known, that in such realistic
gauge theories, as massless QCD or QED, the procedure of the
renormalization is disrupting the initial conformal invariance and
gives rise to an anomaly in the trace of the energy-momentum
tensor~\cite{Crew,ChE}.  Its explicit expression~\cite{ChE}  shows
that the factor $\beta(a)/a$ is the measure of the breaking
of the conformal invariance within the framework of the perturbation
theory expansion in the coupling constant $a$, where in QCD and QED we
will normalize $a$ as $a_s=\alpha_s/\pi$ and $a=\alpha/\pi$
respectively. This means, that in QCD and QED the conformal symmetry
can be effectively restored only in the vicinity of the hypothetical
perturbative fixed points, which satisfy the condition $\beta(a^*)=0$.
In view of this it is rather interesting to understand whether there
are any manifestations of the properties of the initial conformal
symmetry and its violation by the procedure of the renormalization in
the structures of any perturbative series.

By what discussed beyond we will try to convince the readers that the
answers to these questions are positive. We will show that the made
in Ref.[5] careful analysis of the analytical structure of the
perturbative QCD predictions to the certain characteristics of the
$e^+e^-\rightarrow{hadrons}$ and deep-inelastic scattering processes
allowed to reveal the existence of the definite already
proved~\cite{GK} and still non-proved relations between the
coefficients of the perturbative series for the Adler $D$-function of
the {\bf non-singlet} axial currents (or vector currents) and
the polarized Bjorken sum rule (or the Gross-Llewellyn Smith sum
rule). The first quantity was evaluated analytically at the
next-to-leading order (NLO) of perturbative QCD in Ref.[7] (for the
identical result of the independent numerical and analytical
calculations see Ref.[8] and Ref.[9] respectively) and at the
next-to-next-to-leading order (NNLO) in Ref.[10] (for the identical
result of the semi-independent calculations see Ref.[11]). The NLO
coefficient of the perturbative series for the deep-inelastic
scattering sum rules we will be interested in is known from the
results of the analytical calculation of Ref.[12] (later on
independently confirmed in Ref.[13]), while the NNLO corrections to
the polarized Bjorken sum rule and to the Gross-Llewellyn Smith sum
rule were obtained in Ref.[14].

Starting from the beginning of the INR multiloop analytical
calculating project, launched in 1978-1979~\cite{ChT,ChKT},
nobody was  expecting that there are any relations between the
characteristics of the annihilation and deep-inelastic processes.
However, in 1990, when it was already understood by the second and
the third authors of Ref.[10], that the published results of the
analytical calculations of the 4-loop contributions to the QED
$\beta$-function in the $\overline{MS}$-scheme~\cite{GKL1} and to the
Adler $D$-function in QCD~\cite{GKL2} are wrong (for the discussion
see e.g. Ref.[18]), but the 4-loop QED results of Ref.[19] and the
NNLO QCD results of Refs.[10,11,14] were not yet obtained, we were
informed in the quite non-formal form~\cite{Jaffe} about the existence
of the fundamental Crewther relation~\cite{Crew}. This relation is
connecting in the conformal-invariant limit the anomalous 3-point
function of the  axial-vector-vector {\bf
non-singlet} currents with the product of the quark-parton expressions
for the polarized Bjorken sum rule and the $e^+e^-$-annihilation Adler
$D$-function. However, in order to answer to the constructive
question:{\em``What is the status of the Crewther relation in
QCD?''}~\cite{Jaffe} it is  necessary to go beyond the framework of
the conformal-invariant limit and to analyze the structure of the
perturbative QCD corrections to  both sides of the Crewther relation.

Here we will summarize our present understanding of the current status
of the answers to this question using the results of the works
of Refs.[5,6,21] and of the old work of Ref.[22], which became known
to us only recently.

\section{What is the Crewther relation?}

Before starting the presentation of the QCD foundations of Ref.[5] let
us briefly discuss the definite steps of the derivation of the
Crewther relation~\cite{Crew}, repeated also in Ref.[22]. In order
to
complete the analysis of Refs.[3,22], performed in the
$x$-space, we
will follow the studies of Ref.[6] and use the
language of the
momentum space. To our mind, this will allow to
demonstrate more
obviously some basic points, which were not
previously clarified in
Refs.[3,22].

Consider the 3-point
function
\begin{equation}
T_{\mu\alpha\beta}^{abc}(p,q)=i\int<0|TA_{\mu}^{a}(y)V_{\alpha}^{b}(x)
V_{\beta}^{c}(0)|0>e^{ipx+iqy}dxdy=d^{abc}T_{\mu\alpha\beta}(p,q)
\label{1}
\end{equation}
where
$A_{\mu}^{a}(x)=\overline{\psi}\gamma_{\mu}\gamma_{5}(\lambda^{a}/2)\psi$,
$V_{\mu}^{a}(x)=\overline{\psi}\gamma_{\mu}(\lambda^{a}/2)\psi$
are the axial and vector  non-singlet quark currents. The r.h.s.
of Eq.(1) can be expanded in a basis of 3 independent tensor
structures under the condition $(pq)=0$ as
\begin{equation}
\begin{array}{rcl}
T_{\mu\alpha\beta}(p,q) & = &
\xi_1(q^2,p^2)\epsilon_{\mu\alpha\beta\tau}p^{\tau}\\
 & +  &
\xi_2(p^2,q^2)(
q_{\alpha}\epsilon_{\mu\beta\rho\tau}p^{\rho}q^{\tau}-
q_{\beta}\epsilon_{\mu\alpha\rho\tau}p^{\rho}q^{\tau})\\
 & +  &
\xi_3(p^2,q^2)(
p_{\alpha}\epsilon_{\mu\beta\rho\tau}p^{\rho}q^{\tau}+
p_{\beta}\epsilon_{\mu\alpha\rho\tau}p^{\rho}q^{\tau})~~~.
\end{array}
\label{2}
\end{equation}
Taking now the divergency of axial current one can get the following
relation for the invariant amplitude $\xi_1(q^2,p^2)$:
\begin{equation}
q_{\beta}T_{\mu\alpha\beta}(p,q)=\epsilon_{\mu\alpha\rho\tau}
q^{\rho}p^{\tau}\xi_1(q^2,p^2)
\label{3}
\end{equation}
while the property of the conservation of the vector currents implies
that
\begin{equation}
lim|_{p^2\rightarrow{\infty}}p^2\xi_3(q^2,p^2)=-\xi_{1}(q^2,p^2)
\label{4}
\end{equation}
(see Ref.[23] for the discussions of the details of the derivation
of Eqs.(2)-(4)).

In order to clarify the meaning of the second invariant amplitude,
namely $\xi_2(q^2,p^2)$, let us first define the characteristics of
the deep-inelastic processes, namely the polarized Bjorken sum rule
\begin{equation}
Bjp(Q^2)=\int_0^1[g_1^{ep}(x,Q^2)-g_1^{en}(x,Q^2)]dx=\frac{1}{6}
|\frac{g_A}{g_V}|C_{Bjp}(a_s)
\label{5}
\end{equation}
and the Gross-Llewellyn Smith sum rule
\begin{equation}
GLS(Q^2)=\frac{1}{2}\int_0^{1}F_3^{\nu
p+\overline{\nu}p}(x,Q^2)dx= 3C_{GLS}(a_s)~~~~.  \label{6}
\end{equation} The coefficient function $C_{Bjp}(a_s)$ can be found
from the operator-product expansion of two non-singlet vector
currents ~\cite{GL}
\begin{equation}
i\int T{V_{\alpha}^{a}(x)V_{\beta}^{b}(0)}e^{ipx}dx|_{p^2\rightarrow{\infty}}
\approx C_{\alpha\beta\rho}^{P,abc}A_{\rho}^{c}(0)+other~structures
\label{7}
\end{equation}
where
\begin{equation}
C_{\alpha\beta\rho}^{P,abc}\sim
id^{abc}\epsilon_{\alpha\beta\rho\sigma}
\frac{p^{\sigma}}{P^2}C_{Bjp}(a_s)~~~.  \label{8} \end{equation}
and $P^2=-p^2$.
In the case of the definition of the coefficient function of the
Gross-Llewellyn Smith sum rule one should consider the
operator-product expansion of the axial and vector  non-singlet
currents
\begin{equation}
i\int T{A_{\mu}^{a}(x)V_{\nu}^{b}(0)}e^{iqx}dx|_{q^2\rightarrow{\infty}}
\approx C_{\mu\nu\alpha}^{V,ab}V_{\alpha}(0)+other~structures
\label{9}
\end{equation}
where
\begin{equation}
C_{\mu\nu\alpha}^{V,ab}\sim
i\delta^{ab}\epsilon_{\mu\nu\alpha\beta}
\frac{q^{\beta}}{Q^2}C_{GLS}(a_s)~~~
\label{10}
\end{equation}
and $Q^2=-q^2$.
The third important quantity, which will enter into our analysis, is
the QCD coefficient function $C_D^{NS}(a_s)$ of the Adler $D$-function
of the  non-singlet axial currents
\begin{equation}
D^{NS}(a_s)=-12\pi^2q^2\frac{d}{dq^2}\Pi_{NS}(q^2)\sim C_D^{NS}(a_s)
\label{11}
\end{equation}
where $\Pi_{NS}(q^2)$ is defined as
\begin{equation}
i\int<0|TA_{\mu}^a(x)A_{\nu}^b(0)|0>e^{iqx}dx=\delta^{ab}(g_{\mu\nu}q^2-
q_{\mu}q_{\nu})\Pi_{NS}(q^2)~~~.
\label{12}
\end{equation}

At this point we will stop with definitions of the basic quantities
and return to the consideration of the 3-point function of Eq.(1).
Following Ref.[3] one can apply to this correlation function an
operator-product expansion in the limit $|p^2|>>|q^2|$,
$p^2\rightarrow \infty$, namely expand first the $T$-product of two
non-singlet vector currents via Eq.(7) and then take the vacuum
expectation value of the $T$-product of two remaining non-singlet
axial currents defined through Eq.(12). It was shown in Ref.[6] that
these studies imply that
\begin{equation}
\xi_2(q^2,p^2)|_{|p^2|\rightarrow\infty}\rightarrow \frac{1}{p^2}
C_{Bjp}(a_s)\Pi_{NS}(a_s)
\label{13}
\end{equation}
and thus
\begin{equation}
q^2\frac{d}{dq^2}\xi_2(q^2,p^2)|_{|p^2|\rightarrow\infty}\rightarrow
\frac{1}{p^2}C_{Bjp}(a_s)C_D^{NS}(a_s)~~~.
\label{14}
\end{equation}
Equations (13),(14) reflect the physical meaning of the invariant
amplitude $\xi_2(q^2,p^2)$ and should be considered together with the
relations for the invariant amplitudes $\xi_1(q^2,p^2)$ (see Eq.(3))
and $\xi_3(q^2,p^2)$ (see Eq.(4)).

On the other hand, it was shown in Ref.[24] that in a conformal
invariant (c-i)  limit the three-index tensor of Eq.(1) is proportional
to the fermion triangle one-loop graph, constructed from the
massless
fermions, namely that
\begin{equation}
T_{\mu\alpha\beta}^{abc}(p,q)|_{c-i}=
d^{abc}K(a_s)\Delta_{\mu\alpha\beta}^{1-loop}(p,q)~~~.
\label{15}
\end{equation}
In other words, in a conformal invariant limit one has
\begin{equation}
\begin{array}{rcl}
\xi_1^{c-i}(q^2,p^2) & = & K(a_s)
\xi_1^{1-loop}(q^2,p^2)~~~,\\
\xi_2^{c-i}(q^2,p^2) & = & K(a_s)
\xi_2^{1-loop}(q^2,p^2)~~~,\\
\xi_3^{c-i}(q^2,p^2) & = & K(a_s)
\xi_3^{1-loop}(q^2,p^2)~~~.
\end{array}
\label{16}
\end{equation}
Moreover, in view of the Adler-Bardeen theorem~\cite{AB}, which is
insuring the invariant amplitude $\xi_1(q^2,p^2)$, related to the
divergency of axial current (see Eq.(3)), from the renormalizability,
one has $K(a_s)=1$.
The 3-loop light-by-light-type scattering graphs, which were
calculated in Ref.[26] and analyzed in
Ref.[27], do not affect this conclusion. Indeed, in the case
of the 3-point function of the {\bf non-singlet}
axial-vector-vector currents they are contributing to the
higher order QED corrections, while the QCD corrections of the
similar origin are appearing only in the 3-point function
with the {\bf singlet} axial current in one of the vertexes,
which will be not discussed here.

In the case of the consideration of the 3-point function of the
non-singlet axial-vector-vector currents the
property $K(a_s)=1$
is allowing to derive the fundamental Crewther relation
\begin{equation}
C_{Bjp}(a_s(Q^2))C_D^{NS}(a_s(Q^2))|_{c-i}=1~~~,
\label{17}
\end{equation}
which should be valid in the conformal
invariant limit in all orders of perturbation theory. The similar
relation is also true for the  coefficient function $C_{GLS}(a_s)$,
defined by  Eqs.(6),(9),(10)~\cite{ACGJ}. Indeed, considering first
the  operator-product expansion of the axial and vector  non-singlet
currents (see Eq.(9),(10)) in the 3-point function of Eq.(1),
taking  the $T$-product of the remaining vector currents and
repeating the  discussed above analysis, one can find that in the
conformal invariant  limit the following identity takes place:
\begin{equation}
C_{GLS}(a_s(Q^2))C_D^{V}(a_s(Q^2))|_{c-i}=1~~~
\label{18}
\end{equation}
where $C_D^{V}(a_s)$ is the coefficient function of the Adler
$D$-function of two vector currents.

\section{The QCD generalization of the Crewther relation }

It is well known, that the calculations of the
perturbative
theory corrections of the Green functions in the
renormalizable
quantum field models face the necessity of the
introduction of the concrete regularization of the  ultraviolet
divergencies and of the subsequent application of the
renormalization procedures, which as already mentioned, are breaking
the conformal symmetry of the massless free theories. The immediate
consequence of the application of this modern perturbative machinery
is the appearance of the renormalization group $\beta$-functions,
which are governing the energy behavior of the running coupling
constants and are responsible for getting out from the scale-invariant
limit. In QCD the $\beta$-function
\begin{equation}
Q^2\frac{da_s}{dQ^2}=\beta(a_s)=-\sum_{i\geq 0}\beta_i a_s^{i+2}
\label{19}
\end{equation}
was analytically calculated at the 3-loop order in Ref.[27] and
confirmed in Ref.[28] in the framework of the dimensional
regularization~\cite{DR} and the class of minimal subtractions
schemes~\cite{MS}. In our further discussions we will be interested in
the expressions of the first two renormalization-scheme invariant
coefficients $\beta_0$ and $\beta_1$, which are expressed through the
Casimir operators $C_A$, $T_fN_f$ and $C_F$ as
\begin{equation}
\begin{array}{rcl}
\beta_0 & = &
\bigg(\frac{11}{3}C_A-\frac{4}{3}T_fN_f\bigg)\frac{1}{4}~~~,\\ \beta_1
& = &
\bigg(\frac{34}{3}C_A^2-\frac{20}{3}C_AT_fN_f-4C_FT_fN_f\bigg)\frac{1}{16}~~~.
\end{array} \label{20} \end{equation} In general, it is possible to
understand that
\begin{itemize}
\item the perturbative expression for
the QCD $\beta$-function does not contain the terms proportional to
$C_F^Ka_s^K$ $(K\geq 1)$ and
\item the deviation from the conformal
invariant limit $\beta(a_s)=0$ is related to the fact that
$T_fN_f\neq 0$ and $C_A\neq 0$.
\end{itemize}
Moreover, in
perturbative series for physical quantities the latter Casimir
operators can appear only starting from the NLO. Therefore, to study
the theoretical consequences of the property of the conformal
symmetry breaking in the massless gauge models it is necessary to
consider the higher-order perturbative theory approximations for the
physical quantities.

In order to get the non-conformal variants of the Crewther relations
of Eqs.(16),(17) it is necessary to consider the NNLO approximations
of the corresponding basic quantities. Due to the existence of the
dimensional regularization~\cite{DR}, class of the minimal
subtractions schemes~\cite{MS} and the developments in the field of
the creation of the multiloop calculating methods~\cite{ChT,meth},
it became possible to obtain the concrete NNLO results for the Adler
function of the {\bf electromagnetic} quark currents \cite{GKL,SS} by
means of the classical symbolic manipulations program
SCHOONSCHIP~\cite{SCH} and for the QCD coefficient functions of the
polarized Bjorken sum rule and the Gross-Llewellyn Smith sum
rule~\cite{LV} with the help of its more computer-educated younger
follower FORM~\cite{FORM}. Another problem, solved in the process
of the NLO calculations of $C_{Bjp}(a_s)$ and $C_{GLS}(a_s)$ of
Ref.[12] and of the NNLO ones of Ref.[14] is the proper definition of
the axial non-singlet current.

Indeed, it is known, that within dimensional regularization the
analog of the $\gamma_5$-matrix  and
therefore the axial  Ward identities are not well defined. The
most straightforward way, which is allowing to restore  the axial
Ward identities presumes the application of the additional finite
renormalization of the axial currents~\cite{True}, which in the
{\bf non-singlet} case has the following form
\begin{equation}
A_{\mu}^{a}\rightarrow Z_5^{NS}A_{\mu}^a=\tilde{A}_{\mu}^{a}~~~.
\label{21}
\end{equation}
In the $\overline{MS}$-scheme the expression for $Z_5^{NS}$
was calculated at the NLO level in Ref.[12] and
NNLO level in Ref.[14].
It is possible to show, that this additional finite renormalization
allows one to restore the validity of the Adler-Bardeen theorem in
QCD for the anomalous triangle diagram of the axail-vector-vector
{\bf non-singlet} currents (for the related discussions see
Refs.[36-38]). Moreover, using Eq.(21) one can find, that
the NNLO approximation of the QCD coefficient function of the Adler
$D$-function of the {\bf electromagnetic} quark currents
$C_D^{EM}(a_s)$ differs from the NNLO approximation of
$C_D^{NS}(a_s)$ only by the 4-loop light-by-light type diagrams,
which have the {\bf singlet} nature:
\begin{equation}
C_D^{EM}(a_s)=C_D^{NS}(a_s)+C_4^{SI}(a_s)~~~.
\label{23}
\end{equation}
The last scheme-independent contribution has the following form
\cite{GKL,SS}
\begin{equation}
C_4^{SI}(a_s)=\frac{(\sum Q_f)^2}{n_c\sum
Q_f^2}\bigg(\frac{11}{192}-\frac{1}{8}\zeta(3)\bigg)d^{abc}d^{abc}a_s^3
\label{24}
\end{equation}
where $n_c=3$ is the number of colours.
In the $\overline{MS}$-scheme the NNLO analytical expression of
$C_D^{NS}(a_s)$ is known from the results of calculations of
Refs.[10,11]:
$$
C_D^{NS}(a_s) =
1+\frac{3}{4}C_Fa_s
 +  \bigg[-\frac{3}{32}C_F^2
+\bigg(\frac{123}{32}-\frac{11}{4}\zeta(3)\bigg)C_FC_A
+\bigg(-\frac{11}{8}+\zeta(3)\bigg)C_FT_fN_f\bigg]a_s^2  \nonumber
$$
\begin{equation}
 +  \bigg[-\frac{69}{128}C_F^3+\bigg(-\frac{127}{64}-\frac{143}{16}\zeta(3)
+\frac{55}{4}\zeta(5)\bigg)C_F^2C_A
\label{cd21}
\end{equation}
$$
+\bigg(\frac{90445}{3456}
-\frac{2737}{144}\zeta(3)-\frac{55}{24}\zeta(5)\bigg)C_FC_A^2
+\bigg(-\frac{29}{64}+\frac{19}{4}\zeta(3)-5\zeta(5)\bigg)C_F^2T_fN_f
\nonumber
$$
$$
+\bigg(-\frac{485}{27}+\frac{112}{9}\zeta(3)
+\frac{5}{6}\zeta(5)\bigg)C_FC_AT_fN_f
+\bigg(\frac{151}{54}-\frac{19}{9}\zeta(3)\bigg)C_FT_f^2N_f^2\bigg]a_s^3
\nonumber
$$
The similar $\overline{MS}$-scheme result for the coefficient
function $C_{Bjp}(a_s)$ was obtained in Ref.[14] and reads
$$
C_{Bjp}(a_s)=1-\frac{3}{4}C_Fa_s+\bigg[\frac{21}{32}C_F^2-\frac{23}{16}C_FC_A
+\frac{1}{2}C_FT_fN_f\bigg]a_s^2
$$
\begin{equation}
+\bigg[-\frac{3}{128}C_F^3+\bigg(\frac{1241}{576}
-\frac{11}{12}\zeta(3)\bigg)C_F^2C_A
+\bigg(-\frac{5437}{864}+\frac{55}{24}\zeta(5)\bigg)C_FC_A^2
\label{cd22}
\end{equation}
$$
+\bigg(-\frac{133}{576}-\frac{5}{12}\zeta(3)\bigg)C_F^2T_fN_f
+\bigg(\frac{3535}{864}+\frac{3}{4}\zeta(3)-\frac{5}{6}\zeta(5)\bigg)
C_FC_AT_fN_f-\frac{115}{216}C_FT_f^2N_f^2\bigg]a_s^3
$$
while the NNLO coefficient function $C_{GLS}(a_s)$ recieves the
additional NNLO {\bf singlet}-type contribution~\cite{LV}
\begin{equation}
C_{GLS}(a_s)=C_{Bjp}(a_s)+C_{GLS}^{SI}(a_s)
\label{27}
\end{equation}
where
\begin{equation}
C_{GLS}^{SI}(a_s)=\frac{N_f}{n_c}
\bigg(-\frac{11}{192}+\frac{1}{8}\zeta(3)\bigg)d^{abc}d^{abc}
a_s^3~~~.
\label{28}
\end{equation}

From the first superface glance to the pairs of the results of
Eqs.(24),(25) (or to the related ones of Eqs.(22),(26)) one can think
that there is nothing in common between the beautiful, but rather
complicated expressions for the NNLO approximations of the
characteristics of the annihilation processes (see Eqs.(22)-(24)) and
the ones of the deep-inelastic processes (see Eqs.(25)-(27)).
However, after constructing the NNLO QCD variants of the Crewther
formulae of Eqs.(17),(18) and looking to these products more
carefully the following relation was found~\cite{BK}:
\newpage
$$
C_{Bjp}(a_s(Q^2))C_D^{NS}(a_s(Q^2))
$$
\begin{equation}
=1+\frac{\beta^{(2)}(a_s)}{a_s}\bigg[S_1C_Fa_s+\bigg(S_2T_fN_f
+S_3C_A+S_4C_F\bigg)C_Fa_s^2
\bigg]+O(a_s^4)
\label{30}
\end{equation}
where $\beta^{(2)}(a_s)=-\beta_0a_s^2-\beta_1a_s^3$ is the 2-loop
expression for the QCD $\beta$-function with $\beta_0$ and $\beta_1$
defined in Eq.(20) and $S_1$, $S_2$, $S_3$ and $S_4$ are the
analytical numbers, which contain the transcendental Riemann
$\zeta$-functions:
\begin{equation}
\begin{array}{rcl}
S_1 & = &
-\frac{21}{8}+3\zeta(3)~~~,\\
S_2 & = &
\frac{163}{24}-\frac{19}{3}\zeta(3)~~~,\\
S_3 & = & -\frac{629}{32}+\frac{221}{12}\zeta(3)~~~,\\
S_4 & = & \frac{397}{96}+\frac{17}{2}\zeta(3)-15\zeta(5)~~~.
\end{array}
\label{31a}
\end{equation}
The transformation to the non-abelian case of QED can be made by
taking $C_F=1$, $C_A=0$ and $T_fN_f=N$ where $N$ is the number of
massless leptons with identical charges.

The authors of Ref.[5] noticed in Eqs.(28),(29) the following
``seven wonders'' of the generalized Crewther relation, or to be more
precise, of the NNLO non-conformal descrepancy $\Delta_{n-c}=
(C_{D}^{NS}(a_s(Q^2))C_{Bjp}(a_s(Q^2))-1)$ from the
conformal-invariant relations of Eqs.(17),(18):

1. The leading order terms cancels in $\Delta_{n-c}$, which
thus do not contain any $C_Fa_s$-corrections.

2. The NLO corrections give no $C_F^2a_s^2$ terms in $\Delta_{n-c}$.

3. The NNLO corrections give no $C_F^3a_s^3$ terms in $\Delta_{n-c}$.
These three foundations have led to the observation that in the
so-called quenched limit the zero-fermion-loop abelian
terms in the NLO and NNLO approximations of the polarized Bjorken sum
rule and the Gross-Llewellyn Smith sum rule ( explicitely calculated
in Ref.[12] and Ref.[14] respectivley) can be obtained by inversing
the expression of the
one-fermion-loop QED contribution to $C_D^{NS}(a)$, which is related
to the scheme-independent Baker-Johnson QED $F_1$-function,
calculated at the 3-loop level in Ref.[39] (this result was confirmed
in Ref.[40]) and at the 4-loop level in Ref.[19]:
$$
C_{Bjp}(a)|_{quenched~
QED}=1-\frac{3}{4}a+\frac{21}{32}a^2-\frac{31}{128}a^3
$$
\begin{equation}
=\frac{\frac{1}{3}a}{\frac{1}{3}a+\frac{1}{4}a^2-\frac{1}{32}a^3-
\frac{23}{128}a^4}
\label{32a}
\end{equation}
where the factor in the numerator is defined by the 1-loop
contribution to the $F_1$-function.

4. The NNLO light-by-light terms of Eqs.(23),(27) are cancelling
in $\Delta_{n-c}$ (taking equal quark charges in Eq.(23) to obtain
the contribution to $C_D^{V}(a_s)$).

5. The NLO corrections give $C_FC_Aa_s^2$ and $C_FT_fN_f$ terms in
$\Delta_{n-c}$ that are the same ratio as the $C_A$ and $T_fN_f$
terms in $\beta_0$. This reveals the exact factorization of the
$\beta_0a_s^2$-contribution in the NLO correction to
$\Delta_{n-c}$.

6. The NNLO corrections to $\Delta_{n-c}$ can be expressed as the
sum of the terms, proportional to $\beta_0a_s^3$ and $\beta_1a_s^3$
without introduction of the $\beta_0^2a_s^3$-terms, which can be
expected from the simple power-counting arguments.

7. At the NNLO level the $\beta_1a_s^3$ contribution to $\Delta_{n-c}$
occurs with the {\bf same} coefficient that multiplies $\beta_0a_s^2$
term at the NLO.

Before proceeding our discussions let us make several comments,
related to the first three ``wonders'' mentioned above. It can be
shown~\cite{GK} that these three properties are the consequence of
the Adler-Bardeen theorem and of the discussed in Sec.2 initial
Crewther relation~\cite{Crew}, which is valid in the conformal
invariant limit $T_fN_f=0$, $C_A=0$. In fact they were already
effectively discovered in Ref.[22], where the validity of the
3-loop analog of Eq.(30) was conjectured. The result of the NLO
calculations of $C_{Bjp}(a_s)$ (see Ref.[12]), taken in the limit
$C_F=1$, $T_fN_f=0$, $C_A=0$, is confirming the prediction of
Ref.[22] for the abelian zero-fermion-loop NLO contribution to this
sum rule, while the results of the NNLO calculations of the
$F_1$-function~\cite{GKLS} and $C_{Bjp}(a_s)$~\cite{LV} are extending
the obtained in Ref.[22] 3-loop relation to the 4-loop level.
Moreover, it is also possible to understand~\cite{GK}, that the
properties of the cancellation of the $C_F^Ka_s^K$-terms ($1\leq K\leq
3$) in the discovered in Ref.[5] NNLO analog of the Crewther relation
(see Eqs.(28),(29)) can be generalized to the arbitrary order $K$ of
the perturbation theory.

It is worth reminding here the physical meaning of the
Baker-Johnson QED $F_1$-function. In the process of the study of the
finite QED program it was proved that if the condition $F_1(a_*)=0$
takes place, than the property of the finiteness of QED can be
realized in the vicinity of the point $a_*$~\cite{JB,A}. Thus, the
explicite calculations of the coefficients of the perturbative series
for the $F_1$-function can be considered as the important
``experimental'' ingredients of the analysis of the
question of the existence (or, to be more precise,
non-existence) of the perturbative ultaviolet fixed point of the
massless QED.
It is interesting to mention, that the 3-loop analytical calculations
of Ref.[39] revealed the cancellation of the $\zeta(3)$-functions,
which do appear at the intermediate stages of these calculations.
As was mentioned later on in Ref.[40] without any reference, the
fact that the 3-loop coefficient of the $F_1$-function turned out to
be rational can be related to the property of the conformal symmetry
of the massless QED in the quenched approximation.

It was also pointed out in the process of the discusions of the
4-loop results of the 1987-year calculation of the
$F_1$-function~\cite{GKL1}, that it is rather doubtful, that the
result of Ref.[16] (which was found later on to be in error)
contained the $\zeta(5)$-term~\cite{Bender}. Now we know, that in
spite of the appearence of the $\zeta(3)$, $\zeta(4)$ and
$\zeta(5)$-terms at the intermediate stages of calculations of the
4-loop corrections to the $F_1$-function, they are indeed cancelling
out in the ultimate correct result of Ref.[19]. Moreover, due to the
appearence of the personal understanding of the place of the
conformal symmetry in the derivation of the Crewther
relation~\cite{Crew,ACGJ,GK}, we can agree with the comment of
Ref.[40], that the perturbative theory expression for the
$F_1$-function is constrained by the conformal invariant limit of
the massless QED. We can also generalize the proposed in Ref.[40]
hypothesis to arbitrary orders of perturbation theory and make the
conjecture, that it is rather natural to expect that all coefficients
of the $F_1$-function can be rational numbers and will not contain
Riemann $\zeta$-functions. The arguments in favour of the yet
non-proved in detail relations between the appearence of the Riemann
$\zeta$-functions in other results of multiloop QED
calculations, say the 2-loop and 3-loop analytical results
for the anomalous magnetic moment of electron $a_e$ (see Ref.[44] and
Ref.[45] correspondingly) and the theory of knots are given in the
works of Ref.[46].

Let us now return to the discussions of the other four ``wonders'' of
Eq.(28), which were  discovered in Ref.[5]. In fact, since it is
possible to understand (see Ref.[22] and the discussions in Sec.2)
that
\begin{equation}
C_{Bjp}(a_s(Q^2))C_D^{NS}(a_s(Q^2))
=
C_{GLS}(a_s(Q^2))C_D^V(a_s(Q^2))
\label{31}
\end{equation}
there is no place for the {\bf singlet}-type contributions to the QCD
generalization of the Crewther relation for the 3-point function of
the axial-vector-vector {\bf non-singlet} currents. The mentioned in
the item 4 cancellation of the light-by-light-type order
$\alpha_s^3$-diagarms, which produce the colour structure
$(d^{abc})^2$, is nothing more than the consequence of Eq.(31). The
same equation is insuring this product from the contribution of the
3-loop triangle-type {\bf singlet} diagrams, calculated in Ref.[47]
and re-calculated with the same result in the works of Ref.[48]
(for the details see Ref.[49]).

Three last ``wonders'' are leading to the striking indications on
the factorization of the term $(\beta^{(2)}(a_s)/a_s)$ in the NNLO
generalization of the Crewther relation of Eq.(28).
It should be stressed, that we prefer to think about the theoretical
origine of the factorization of this special term, but not simply of
the 2-loop approximation of the QCD $\beta$-function
$\beta^{(2)}(a_s)$, since in case of the cancellation of
of $a_s$ in the denominator of the conformal symmetry breaking
term $(\beta^{(2)}(a_s)/a_s)$ by the extra power of $a_s$ in the
square bracket of the r.h.s. of Eq.(28), we will get in it free
Casimir operator $C_F$, which is not multiplied by the coupling
constant $a_s$. In its turn, this will contradict the general
property of the gauge structure of the concrete expressions for the
perturbative contributions, which presumes the appearence of the
Casimir operators only in the combinations, proportional to
$C_Fa_s$, $C_F^2a_s^2$, $C_FC_Aa_s^2$, $C_FT_fN_fa_s^2$, etc.

Another yet non-proved property of the QCD generalization of the
Crewther relation is following from the ``wonder'' N6. Indeed, the
absence of the NNLO contribution to $\Delta_{n-c}$ of the terms,
proportional to $\beta_0^2a_s^3$ is giving us the idea, that the
explicite expression of the $\Delta_{n-c}$-factor can not contain
the contribution, proportional to $(\beta^{(2)}(a_s)/a_s)^2$.
Though we cannot yet be sure, that the factor $(\beta(a_s)/a_s)$ can
be factored out of $\Delta_{n-c}$ beyond the NNLO, it seems to the
authors of Ref.[5] most likely, that at any given order
$a_s^K~(K\geq 1)$, one will encounter in $\Delta_{n-c}$ only the
coefficients $\beta_i|i<K$, multiplied by linear combinations of
colour factors.

In fact we think, that the hypothesis $\Delta_{n-c}\sim
(\beta(a_s)/a_s)$ merits close attention. In Ref.[6] the attempt to
study the theoretical consequencies of this hypothesis in more detail
was made. Using the assumption, that the factorized factor, which is
appearing in the r.h.s. of Eq.(28) has the origin, similar to the one
of the measure of the conformal invariance breaking within
perturbation theory framework $(\beta(a_s)/a_s)$ in the explicite
expression for the anomaly of the energy momentum
tensor~\cite{ChE}, it was proposed in Ref.[6] to rewrite Eqs.(16)
for the QCD expressions of the tensor structures of the 3-point
function of Eq.(2) as
\begin{equation}
\begin{array}{rcl}
\xi_1(q^2,p^2) & = & K(a_s)
\xi_1^{1-loop}(q^2,p^2)~~~,\\
\xi_2(q^2,p^2) & = & \bigg(K(a_s)+\frac{\beta(a_s)}{a_s}v_2(q^2,p^2,a_s)\bigg)
\xi_2^{1-loop}(q^2,p^2)~~~,\\
\xi_3(q^2,p^2) & = &
\bigg(K(a_s)+\frac{\beta(a_s)}{a_s}v_3(q^2,p^2,a_s)\bigg)
\xi_3^{1-loop}(q^2,p^2)~~~.
\end{array}
\label{32}
\end{equation}
where $v_2$ and $v_3$ are dimensionless functions, which can be
constrained from the identity
\begin{equation}
q^2\frac{d}{dq^2}\xi_2(q^2,p^2)=-p^2\frac{d}{dq^2}\xi_3(q^2,p^2)
-\xi_2(q^2,p^2)~~~.
\label{33}
\end{equation}
This equation can be obtained from the derived in Ref.[23] Ward
identity
\begin{equation}
-\xi_1(q^2,p^2)=q^2\xi_2(q^2,p^2)+p^2\xi_3(q^2,p^2)
\label{34}
\end{equation}
after noting that according to the Adler-Bardeen theorem the
function $\xi_1(q^2,p^2)$ is simply the unrenormalizable number.
Keeping in mind Eqs.(32), which were written down after using the
{\bf still non-proved} assumptions that the factor $(\beta(a_s)/a_s)$
is
(a) indeed factorized in the QCD generalization of the Crewther
relation and
(b) it can really manifest itself in the perturbative expressions of
$\xi_2(q^2,p^2)$ and $\xi_3(q^2,p^2)$ in the form, suggested in
Eqs.(32)
it is possible to show, that in QCD the Crewther product of Eq.(17)
takes the following form~\cite{GK}:
\begin{equation}
C_{Bjp}(a_s(Q^2))C_D^{NS}(a_s(Q^2))=1+\frac{\beta(a_s)}{a_s}r(a_s)
\label{35}
\end{equation}
where $r(a_s)$ is a polynomial in powers of $a_s$, which is not fixed
in the approach of Ref.[6]. Therefore, the non-proved assumptions (a)
and (b) are closely related.

Of course, a lot of work should be still done in order to find the
really proved theoretical support in favour of these assumptions.
Indeed, it is still necessary to understand on the diagrammatic
language the origine of the appearence of the term, proportional
to $(\beta^{(2)}(a_s)/a_s)$ in the discovered in Ref.[5] QCD
generalization of the Crewther relation (see Eq.(28)). In view of the
presented in Ref.[6] considerations we think, that the analysis of
this problem should be started from the explicite calculations of the
NLO perturbative QCD corrections to the invariant amplitude
$\xi_2(q^2,p^2)$ of the 3-point function of Eq.(1). The second
non-solved question is related to the necessity of the understanding,
whether the factor $(\beta(a_s)/a_s)$, presumably related to the
property of the conformal symmetry breaking in massless QCD, is
indeed factorized in the generalized Crewther relation in all orders
of perturbation theory. We believe, that the attraction of the
formalizm of the non-conformal Ward identities, developed in
Ref.[50], can be rather useful for the study of this interesting
problem.

\section{The generalzied Crewther relation and the ``commensurate
scale relations''}

It should be stressed that the QCD generalization of the Crewther
relation of Eq.(28) was discovered in Ref.[5] using the NNLO
$\overline{MS}$-scheme results for $C_D^{NS}(a_s)$~\cite{GKL,SS}
and the similar ones for $C_{Bjp}(a_s)$~\cite{LV}. However, it is
known that on the contrary to the QED on-shell scheme, which is
distinguished by the kinematics of the experimental measurements
of the magnetic moments of electron and muon, the
$\overline{MS}$-scheme is, regorously speaking, not physical one and
is chosen as the reference scheme in the QCD phenomenology only by the
convention between theoreticians and
experimentalists. In view of this the number of approaches of dealing
with the existing problem of fixing the scheme-dependence
uncertainties in QCD was proposed. Among these methods are the
principle of minimal sensitivety~\cite{PMS}, the effective charges
approach~\cite{Gru,KKP} (which is known to be {\em a posteriori}
equivalent to the scheme-invariant perturbation theory, developed in
Refs.[54,55]) and the BLM approach~\cite{BLM}, generalized to the
NNLO level in Ref.[57] (see also Ref.[58]).

Following the work of Ref.[21] we will show that it is possible to
apply the methods of Refs.[51-56] in order to rewrite the generalized
Crewther relation of Eq.(28) in the different form and to derive the
variants of the obtained in Ref.[59] so-called ``commensurate scale
relations''. Let us first define the effective charges of the
coefficient functions $C_D^{NS}(a_s)$ and $C_{Bjp}(a_s)$ as
\begin{equation}
\begin{array}{rcl}
C_D^{NS}(a_s) & = & 1+\hat{a}_D(Q^2/\Lambda_D^2)~~~, \\
C_{Bjp}(a_s) & = & 1-\hat{a}_{Bjp}(Q^2/\Lambda_{Bjp}^2)~~~,
\end{array}
\label{36}
\end{equation}
where
$\hat{a}_D=\frac{3C_F}{4}a_s^D$
and
$\hat{a}_{Bjp}=\frac{3C_F}{4}a_s^{Bjp}$ can be related to
the $\overline{MS}$-scheme results using the following equation
$$
a_s^{Bjp(D)}(Q^2/\Lambda_{Bjp(D)}^2)  =
a_s(Q^2/\Lambda_{\overline{MS}}^2)+
(A_{1(2)}+B_{1(2)}\beta_0)a_s^2(Q^2/\Lambda_{\overline{MS}}^2)
$$
\begin{equation}
+ (C_{1(2)}+D_{1(2)}\beta_0+E_{1(2)}\beta_0^2+B_{1(2)}\beta_1)
a_s^3(Q^2/\Lambda_{\overline{MS}}^2)~~~~.
\label{37}
\end{equation}
Here
$\Lambda_{Bjp(D)}^2=\Lambda_{\overline{MS}}^2exp
\bigg(\frac{A_{1(2)}+B_{1(2)}\beta_0}{\beta_0}\bigg)$ are the
effective scales of $C_{Bjp}(a_s)$ and $C_D^{NS}(a_s)$-coefficient
functions, $\beta_0$ and $\beta_1$ are the scheme-invariant
coefficients of the QCD $\beta$-function (see Eq.(20)) and the exact
expressions of the terms $A_i,B_i,C_i,D_i,E_i~(i=1,2)$ can be
extracted from the NNLO results of Eqs.(25),(24). Two effective
charges $a_s^{Bjp}$ and $a_s^D$ can be related as
\newpage
$$
a_s^{Bjp}(Q^2/\Lambda_{Bjp}^2)  =
a_s^D(Q^2/\Lambda_{D}^2)+
(A_{12}+B_{12}\beta_0)(a_s^D(Q^2/\Lambda_{D}^2)^2 
$$
\begin{equation}
+ (C_{12}+D_{12}\beta_0+E_{12}\beta_0^2+B_{12}\beta_1)
(a_s^D(Q^2/\Lambda_{D}^2))^3~~~~
\label{38}
\end{equation}
where $A_{12}=A_1-A_2$, $B_{12}=B_1-B_2$,
$C_{12}=C_1-C_2-2(A_1-A_2)A_2$,
$D_{12}=D_1-D_2-2(A_1B_2+A_2B_1)+4A_2B_1$,
$E_{12}=E_1-E_2-2(B_1-B_2)B_2$ are the scheme-independent
coefficients.

The basic aim of the applications of the BLM criterion~\cite{BLM} to
the series of Eq.(38) is the construction of the $N_f$-independent
perturbative scheme-independent expansion, which can play the role,
similar to the one of the Baker-Johnson scheme-independent
$F_1$-funcion in QED. At the NNLO level this aim can be achieved with
the help of the developed in Ref.[57] single-scale generalization of
the BLM method, which leads to the following redifinition of the
scale of the $a_s^D$-effective coupling constant from $Q^2$ to
$\overline{Q}^{*2}$~\cite{BGKL}: \\
$
ln\bigg(\overline{Q}^{*2}/Q^2\bigg)=-B_{12}+[\beta_0(B_{12}^2-E_{12})
+2A_{12}B_{12}-D_{12}]a_s^D(\overline{Q}^{*2}/\Lambda_D^2)$.\\
It should be stressed, that this procedure is effectively
eliminating the dependence of the r.h.s. of Eq.(38) from the
$\beta_0$-coefficient. Moreover, since the coefficient before the
$\beta_1$-term in the NNLO contribution to Eq.(38)
can be chosen to be
equal to the
coefficient $B_{12}$ of the NLO correction, the absorption of the
proportional to $N_f$ (and thus to $\beta_0$) NLO term into the scale
$\overline{Q}^{*}$ automatically leads to the nullification of the
proportional to $\beta_1$ NNLO contribution in this relations
bewteen effective charges.

Taking now into account the concrete expressions for the effective
charges $a_s^D$ and $a_s^{Bjp}$ (see Eqs.(24),(25)) one can
find~\cite{BGKL}
$ln\bigg(\overline{Q}^{*2}/Q^2\bigg)
=  \frac{7}{2}-4\zeta(3)+a_s^D(\overline{Q}^*)
\bigg[\bigg(\frac{11}{12}+\frac{56}{3}\zeta(3)-16\zeta^2(3)\bigg)\beta_0
+
\frac{13}{18}C_A-\frac{2}{3}C_A\zeta(3)-\frac{145}{72}C_F
-\frac{46}{3}\zeta(3)C_F+20\zeta(5)C_F\bigg]$
and
\begin{equation}
\hat{a}_{Bjp}(Q)=\hat{a}_D(\overline{Q}^*)-\hat{a}_D^2(\overline{Q}^*)
+\hat{a}_D^3(\overline{Q}^*)+...~~~.
\label{41}
\end{equation}
Equation (39) is representing the example of the single-scale variant of the
``commensurate scale relation'' of Ref.[59]. It is the
consequance of the following variant of the generalized Crewther
relation of Eq.(28):
\begin{equation}
C_{Bjp}(a_s^{Bjp}(Q^2))C_D^{NS}(a_s^D(\overline{Q}^{*2})=1
\label{42}
\end{equation}
where the non-conformal term $\Delta_{n-c}$ in Eq.(28) is absorbed
into the scale of the coefficient function $C_D^{NS}$. It should be
stressed, that the property of the factorization of the term
$(\beta^{(2)}(a_s)/a_s)$ in the expression of Eq.(28) for
$\Delta_{n-c}$, discovered in Ref.[5], turns out to be the necessary
and the sufficient condition, which allowed the authors of Ref.[21]
to rewrite the generalized Crewther relation of Eq.(28) in the
form of the geometric progression
of Eq.(39).  Note also, that the less convenient for practical
applications multi-scale variant of the ``commensurate scale
relation''of Eq.(39), which was previously derived in Ref.[59],
is also the consequence of the definite variant of the generalized
Crewther relation of Eq.(28).

\section{The generalized Crewther relation and the experiment}

After reading the previous Sections one can be
interested in getting the understanding whether it is possible to
build the bridge between the pure theoretical studies presented
above and the existing experimental data for the characteristics of
the $e^+e^-$-annihilation and deep-inelastic processes, which enter
in the Crewther relation and its QCD generalizations.
The first attempts to analyze this problem was made in Ref.[21] and
in the closely related work of Ref.[58].
In this Section we will follow the original considerations of
Ref.[21], leaving the comments to the outcomes of Ref.[58] for the
possible future presentation.

Let us first mention, that while the deep-inelastic scattering sum
rules are measured in the Eucledian region, the real experimental
information for the basic chracteristic of the $e^+e^-$-annihilation
channel
$R_{e^+e^-}(s)=\sigma_{tot}(e^+e^-\rightarrow{hadrons})
/\sigma(e^+e^-\rightarrow\mu^+\mu^-)$ is coming from the measurements
in the Minkowskian region. If one is sufficiently far from the
resonance production thresholds, it is possible to relate the
perturbative expression for $R_{e^+e^-}(s)$ with the coefficient
function  $C_D^{EM}(Q^2)$ by the following relation
\begin{equation}
R_{e^+e^-}(s)=\frac{3\sum_f Q_f^2}{2\pi
i}\int_{-s-i\epsilon}^{-s+i\epsilon}\frac{d\tau}{\tau}C_D^{EM}(a_s(\tau))~~~.
\label{43}
\end{equation}
In general this procedure results in the appearence of the
$\pi^2$-like terms in the coefficient function of the $R$-ratio
starting from the NNLO-level (see e.g. Ref[60]). In fact they can be
taken into account in the formulae of Eqs.(37),(38) using the
following shifts $E_2\rightarrow E_2-\pi^2/3$, $E_{12}\rightarrow
E_{12}+\pi^2/3$. The higher-order $\pi^2$-contributions to
$R_{e^+e^-}(s)$ were calculated explicitely in Refs.[61,62]. However,
since at the present stage we are interested in the consequencies of
the NNLO generalizations of the Crewther relation, we will not sum-up
these higher-order $\pi^2$-contributions (like it was proposed to do
e.g. in Ref.[63]), but trancate the corresponding perturbative series
in the time-like region at the NNLO level. In order to obtain the
relation between the time-like scale
$\sqrt{s^*}$ of the $R_{e^+e^-}$-ratio and the space-like scales of the
deep-inelastic scattering sum rules it is not only necessary to take
into account the NNLO $\pi^2$-contributions to the $R$-ratio, but to
perform the replacements of the effective charges
$a_s^D(Q^*)\rightarrow a_s^R(\sqrt{s^*})$ as well. Than the resulting
relation reads~\cite{BGKL}:
$$
ln\bigg(\frac{Q^2}{s^*}\bigg)
=  -\frac{7}{2}+4\zeta(3)-a_s^R(s^*)
\bigg[\bigg(\frac{11}{12}+\frac{56}{3}\zeta(3)-16\zeta^2(3)-\frac{\pi^2}{3}
\bigg)\beta_0
$$
\begin{equation}
\frac{13}{18}C_A-\frac{2}{3}C_A\zeta(3)-\frac{145}{72}C_F
-\frac{46}{3}\zeta(3)C_F+20\zeta(5)C_F\bigg]~~~.
\label{44}
\end{equation}
The variants of the generalized Crewther relation  can be
then written down in the following form~\cite{BGKL}
\begin{equation}
\frac{1}{3\sum_f Q_f^2}C_{Bjp}(a_s^{Bjp}(Q^2))R_{e^+e^-}(a_s^R(s^*))
 =  1+C_4^{SI}(a_s^R(s^*))~~~,
\label{45}
\end{equation}
$$
\frac{1}{3\sum_f Q_f^2}C_{GLS}(a_s^{GLS}(Q^2))R_{e^+e^-}(a_s^R(s^*))
=  1+C_4^{SI}(a_s^R(s^*))+C_{GLS}^{SI}(a_s^{GLS}(Q^2))~~~,
$$
where $C_4^{SI}$ abd $C_{GLS}^{SI}$ are the singlet-type
contributions to $C_D^{EM}(a_s)$ and $C_{GLS}(a_s)$ which are defined
in Eq.(23) and Eq.(27) respectively. In fact, since the numerical
values of these contributions are very small, it is reasonable to
neglect them in the phenomenologically oriented discussions and thus
assume, that $C_D^{EM}(a_s)\approx C_D^{NS}(a_s)$ and $C_{Bjp}(a_s)
\approx C_{GLS}(a_s)$.

It should be stressed, that the theoretical expressions of Eq.(43)
are derived in the framework of the perturbation theory and do not
involve the non-perturbative contributions to
$C_D^{EM}(a_s)$~\cite{ShVZ} (and thus $R$-ratio), and to the
deep-inelastic scattering sum rules, theoretically calculated in
Ref.[65]
and numerically estimated using the QCD sum-rules formalizm in
Ref.[66]. In fact it is known, that these contributions are very
important for the analysis of the low-energy experimental data for
the $R$-ratio~\cite{EKV} and the Gross-Llewellyn Smith and the
polarized Bjorken sum rules (see Refs.[68,69] and Refs.[70,71]
respectively).  Therefore, in the estimates, which have the aim to
clarify the experimental status of the perturbative QCD
generalizations of the Crewther relation, it is necessary to chose
the scales in the regions of energies, where the non-perturbative
effects can be safely neglected. For the $e^+e^-$-annihilation
$R$-ratio the lowest energy region,
which is satisfying this criterion, is
$4~GeV\leq\sqrt{s}\leq 8~GeV$.  It is
typical to the case of taking $N_f=4$ numbers af active flavours and
is
lying above the thresholds of
the production of the $c\overline{c}$-bound states $\sqrt{s}\approx
3~GeV$ and beyond the thresholds of the production of
$b\overline{b}$-bound states $\sqrt{s}\approx 10~GeV$. The detailed
QCD analysis of the $e^+e^-$-data for the $R$-ratio in this region was
made previously in the number of the works on the subject (see e.g.
Refs.[72,73]). In Ref.[21] the result of the most recent similar fit
of Ref.[74] was used, which provide the authors of Ref.[21] with the
constarint $(1/3\sum_f Q_f^2)R_{e^+e^-}(\sqrt{s^*}=5.0~GeV)\approx
1.08\pm 0.03$ and thus $a_s^R(\sqrt{s^*}=5.0~GeV)\approx 0.08\pm
0.03$. Equations (42),(43) then imply that the corresponding
estimates for the coefficient functions of the sum rules  lie
on the segment line which connects the following three
values~\cite{BGKL}:
\begin{equation}
\begin{array}{rcl}
C_{Bjp}(Q=11.13~GeV) & \approx  & C_{GLS}(Q=11.13~GeV)\approx
0.952~~,\\
C_{Bjp}(Q=12.33~GeV) & \approx &
C_{GLS}(Q=12.33~GeV) \approx 0.926~~,\\
C_{Bjp}(Q=13.53~GeV) &
\approx & C_{GLS}(Q=13.53~GeV) \approx 0.900~~.
\end{array}
\label{csr}
\end{equation}
The corresponding values of the effective
coupling constants lie in the line segment, which connects the three
related points \begin{equation} \begin{array}{rcl}
a_s^{Bjp}(Q=11.13~GeV) & \approx  & a_s^{GLS}(Q=11.13~GeV) \approx
0.048~~,\\ a_s^{Bjp}(Q=12.33~GeV) & \approx  &
a_s^{GLS}(Q=12.33~GeV) \approx 0.074~~,\\ a_s^{Bjp}(Q=13.53~GeV) & \approx
& a_s^{GLS}(Q=13.53~GeV) \approx 0.1~~.  \end{array} \label{car}
\end{equation}
The appearence of three corresponding values of Eqs.(44) and Eqs.(45)
are related to the uncertainties in the definition of $a_s^R(\sqrt{s}
=5~GeV)$, which are translated by Eq.(42) into three related numbers.
The predictions of Eq.(45) could be tested experimentally.

It should
be stressed, that {\bf at present} the measurements of the polarized
Bjorken sum rule are allowing to obtain its precise
experimental values in the regions of energies of over $3~GeV^2$~(see
Ref.[75]) and $10~GeV^2$ (see Ref.[76]). The recent measurements of
the Gross-Llewellyn Smith sum rule are also preformed at relatively
small values of $Q^2$ (see Ref.[69]). However, for the estimates,
aimed to the comparison with the results of Eqs.(45), it was
proposed in Ref.[21] to use the results of the theoretical
extrapolation~\cite{KS} of the available experimental data of the
CCFR collaboration~\cite{CCFR} to the wide region of energies. The
indirect determination of the values of the Gross-Llewellyn Smith
sum rule for $3~GeV^2\leq Q^2\leq 500~GeV^2$ of Ref.[77]
gives the value
\begin{equation}
a_s^{GLS}(Q=12.25~GeV)\approx 0.093\pm 0.042~~~.
\label{gls1}
\end{equation}
This interval crosses the line of the estimates of Eq.(25). To the
point of view of the authors of Ref.[21] this fact gives empirical
support for the generalized Crewther relation, written down in the
form of Eq.(40).

In order to clarify the meaning of these numbers it seems to us
rather instructive to return to the commonly used language of the
$\overline{MS}$-scheme. Let us first analyse the question of the
extraction of the value of $\Lambda_{\overline{MS}}$ from the
fits of the $e^+e^-$-data of Ref.[74]. The results of the NNLO
calculations of Refs.[10,11] relate the effective charge of the
$e^+e^-$-annihilation to the coupling constant $a_s$ in the
$\overline{MS}$-scheme as $a_s^R=a_s[1+1.524a_s-11.52a_s^2]$.
Here $N_f$=4 is taken and the small contribution of $C_4^{SI}(a_s)$
is neglected. Using this relation, we find that the used by us
experimentally motivated number for $a_s^{R}(\sqrt{s}=5~GeV)$
corresponds to the following value of the coupling constant
$\alpha_s$ in the $\overline{MS}$-scheme: $\alpha_s(\sqrt{s}=5~GeV)
=0.238^{+0.062}_{-0.087}$. It is known that the coupling constant
$\alpha_s$ can be expressed through the QCD scale parameter
$\Lambda_{\overline{MS}}$ as
$$
\frac{\alpha_s}{4\pi}  =
\frac{1}{\beta_0\ln(Q^2/\Lambda_{\overline{MS}}^2)}
-\frac{\beta_1\ln\ln(Q^2/\Lambda_{\overline{MS}}^2)}{\beta_0^2\ln^2(Q^2/
\Lambda_{\overline{MS}}^2)}
$$
\begin{equation}
+ \frac{\beta_1^2\ln^2\ln(Q^2/\Lambda_{\overline{MS}}^2)-
\beta_1^2\ln\ln(Q^2/\Lambda_{\overline{MS}}^2)+\beta_2\beta_0-\beta_1^2}
{\beta_0^5\ln^3(Q^2/\Lambda_{\overline{MS}}^2)}~~~.
\end{equation}
Combining Eq.(47) with the obtained above value of
$\alpha_s(\sqrt{s})$ (normalized to $N_f=4$ numbers of active
flavours) we find the following interval of the values of the
parameter $\Lambda_{\overline{MS}}^{(4)}=410^{+320}_{-330}~MeV$.
Evolving this result through the threshold of the production of the
$b\overline{b}$-bound $M=2m_b\approx 9~GeV$ using the
approximate formula of Ref.[79] we get
$\Lambda_{\overline{MS}}^{(5)}=275^{+246}_{-230}~MeV$ and
\begin{equation}
\alpha_s(M_Z)=0.123^{+0.014}_{-0.027}~~~.
\label{alphas}
\end{equation}
This value is compatible with the result $\alpha_s(M_Z)=0.124\pm
0.021$, which comes from the recent detailed fit of the available
data in $e^+e^-$-annihilation from $\sqrt{s}=20~GeV$ tp $65~GeV$
\cite{fit} (for the detailed discussions of this result see the
recent review of Ref.[81]. Thus the current measurements of
$R_{e^+e^-}$ suffer from sizable experimental uncertainties.

The similar situation also holds in the case of the existing
experimentally motivated data for the Gross-Llewellyn Smith
sum rule at the energies, higher than $Q^2>5~GeV^2$ (see, for
example, the results of the recent analysis of Ref.[69]). In order to
demonstrate explicitely that the used outcomes of the
extrapolation-extraction of the Gross-Llewellyn Smith sum rule value
at $Q^2=150~GeV^2$ (see Ref.[77]) has definite theoretical
and experimental uncertainties, we transform the results of Eq.(46)
into the $\overline{MS}$-scheme and obtain
$\alpha_s(Q=12.25~GeV)\approx 0.220^{+0.078}_{-0.088}$.
This estimate might be  weekly senistive to the change of $N_f$ from
$N_f=4$ (which is typical to the fits of the deep-inelastic data)
to $N_f=5$, which is more appropriate to the scale $Q^2=150~GeV^2$.
Therefore, we will
extract from the high-energy  results of Ref.[77] the value of
$\Lambda_{\overline{MS}}^{(5)}$ and will get
\begin{equation}
\alpha_s(M_Z)=0.119^{+0.010}_{-0.018}~~~.
\label{alphas1}
\end{equation}
One can see that this estimate has larger uncertainties, than the
value $\alpha_s(M_Z)=0.109\pm 0.003(stat)\pm 0.005(syst)\pm
0.003 (thor)$, recently extracted in Ref.[82] from the analysis
of the CCFR data for $xF_3$ structure function using the information
about the NNLO corrections to the coefficient functions~\cite{ZV2}
and the results of the recent  calculations of the NNLO
corrections to the anomalous dimensions of the non-singlet
operators~\cite{new}. Therefore, the current experimentally-motivated
estimate of the value of the Gross-Llewellyn Smith sum rule at
$Q^2=150~GeV^2$ (see Ref.[77]) also suffers from the definite
theoretical and experimental uncertainties.

These discussions are giving the additional arguments in favour
of the physical conclusions of Ref.[21]: in order to check the
consequencies of the perturbative generalizations of the Crewther relation at
more high confidence level it can be rather helpul, first, to reduce
the experimental error of the measurements of $R_{e^+e^-}$ at $\sqrt{s}\approx
5~GeV$ and, secondly, to have more precise information on the value of
the Gross-Llewellyn Smith sum rule or of the Bjorken polarized sum rule
at $Q^2\approx 150~GeV^2$. The first problem can be attacked after
starting the operation of the $c-\tau$-factory, while the possible future
study of the deep-inelastic scattering with both polarized electron and
proton beams can open the window for direct measurements of the
polarized Bjorken sum rule at high momentum transfer~\cite{HERA}.

Another interesting, to our point of view, proposal
is to try to measure
the value of the polarized Bjorken sum rule (or the Gross-Llewellyn Smith
sum rules) at the scale $Q^2=25~GeV^2$. It can be useful for the study of the
status of the perturbative generalization of the  Crewther relation, written
down in the form of Eq.(28), derived in Ref.[5]. We hope to
return to the more detailed analysis of this problem in our
possible future  work.

\section{Is there any  generalzation of the Crewther relation for
the 3-point function of singlet axial-non-singlet vector-vector
currents? }

As was already explained in Sec.2, the classical Crewther relation
was originally derived in Ref.[3] for the 3-point function of the
axial-vector-vector non-singlet currents. For the sake of completeness let
us now discuss the case of the analogouse 3-point function with the axial
singlet current :
\begin{equation}
T_{\mu\alpha\beta}^{ab}(p,q)=i\int<0|TA_{\mu}(y)
V_{\alpha}^a(x)V_{\beta}^b(0)|0>e^{ipx+iqy}dxdy
\label{avv}
\end{equation}
where $A_{\mu}=\overline{\psi}\gamma_{\mu}\gamma_{5}\psi$.
In this case it is also possible to try to repeat the considerations
of Sec.2 and to think about the possibility of the derivation
of the Crewther-type relations.
Indeed,
keeping the singlet structure in the operator-product expansion of
the two non-singlet vector currents, one can get
\begin{equation}
i\int T V_{\alpha}^aV_{\beta}^be^{ipx}dx|_{p^2\rightarrow\infty}
\approx C_{\alpha\beta\rho}^{SI,ab}A_{\rho}(o)+other~ structures~~~.
\label{vv}
\end{equation}
In analogy with Eq.(10) one finds that
\begin{equation}
C_{\mu\nu\alpha}^{SI,ab}\sim i\delta^{ab}\epsilon_{\mu\nu\alpha\beta}
\frac{q^{\beta}}{Q^2}C_{EJ}^{SI}(a_s)~~~~.
\label{csi}
\end{equation}
The coefficient function $C_{EJ}^{SI}(a_s)$ is entering in the
definition of the so-called Ellis-Jaffe sum rule as~\cite{Larin2}
\begin{equation}
\begin{array}{rcl}
EJ(Q^2) & =  & \int_0^1 g_1^{p(n)}(x,Q^2) =\bigg(\pm\frac{1}{12}|g_A|
+\frac{1}{36}a_8\bigg)C_{EJ}^{NS}(a_s)\\
&+&
\frac{1}{9}\Delta\Sigma(\mu^2)\times AD\times C_{EJ}^{SI}(a_s)
\end{array}
\label{csra}
\end{equation}
where $C_{EJ}^{NS}(a_s)=C_{Bjp}(a_s)$,
$AD=exp\bigg(\int_{a_s(\mu^2)}^{a_s(Q^2)}\frac{\gamma^{SI}(x)}
{\beta(x)}dx\bigg)$ and $|g_A|=\Delta  u-\Delta d$, $a_8=\Delta u+
\Delta d-2\Delta s$, $\Delta\Sigma=\Delta u+\Delta d+\Delta s$,
$\Delta u$, $\Delta d$, $\Delta s$ can be interpreted as the measure
of the polarization of quarks in a nucleon,
$C_{EJ}^{NS}(a_s)=C_{Bjp}(a_s)$ and $\gamma^{SI}(a_s)$ is the
anomalous dimension of the {\bf singlet} axial current, which can be
calculated within the $\overline{MS}$-scheme provided the additional
finite renormalization of the singlet axial current is made:
\begin{equation}
A_{\mu}\rightarrow Z_5^{SI}A_{\mu}=\tilde{A}_{\mu}
\label{axialred}
\end{equation}
It was shown in Ref.[38] that the finite renormalization constant
$Z_5^{SI}$ is different from the defined in Eq.(21) ``non-singlet
axial charge'' due to the additional contribution to $Z_5^{SI}$ of
the 3-loop light-by-light-type triangle diagrams. At the 2-loop order
this difference is $Z_5^{SI}=Z_5^{NS}+\frac{3}{32}C_FN_fa_s^2$~\cite{Larin1}.
The similar difference is also appearing in the coefficient function
$C_D^{SI}(a_s)$ for the Adler $D$-function of two singlet currents, which
can be constructed from the following correlation function
\begin{equation}
i\int<0|T A_{\mu}(x) A_{\nu}(0)|0>e^{iqx}dx=\Pi_{\mu\nu}^{SI}(q^2)~~~.
\label{ax3}
\end{equation}
Staring from the 3-loop order the coefficient function $C_D^{SI}(a_s)$ differs
from the coefficient function $C_D^{NS}(a_s)$ by the light-by-light-type
diagrams, first calculated in Ref.[47] in the limit of the heavy top-quark
mass $m_t$:
\begin{equation}
C_D^{SI}(a_s)=C_D^{NS}(a_s)+\Delta C_D^{SI}(a_s)~~~.
\label{new2}
\end{equation}
In the case of the lighter quarks $m_q<<m_t$ the result for the 3-loop
light-by-light-type contribution can be extracted from the first paper
of Ref.[49].

Let us now return to the discussion of the singlet contribution 
to the Ellis-Jaffe sum rule. The NLO corrections to the coefficient 
function $C_{EJ}^{SI}(a_s)$ was calculated in Ref.[86]. It is 
different from the result of the calculation of Ref.[12] due to the 
appearence of the light-by-light-type singlet diagrams
\begin{equation}
C_{EJ}^{SI}(a_s)=C_{EJ}^{NS}(a_s)+\Delta C_{EJ}^{SI}(a_s)
=C_{Bjp}(a_s)+\Delta C_{EJ}^{SI}(a_s)
\label{57}
\end{equation}
where at the NLO level $\Delta 
C_{EJ}^{SI}(a_s)=(\zeta(3)+\frac{1}{24}C_FT_fN_f)a_s^2$~\cite{Larin2}.
This result is in agreement with the calculation of Ref.[83]. The 
anomalous dimension $\gamma^{SI}(a_s)$ was calculated in the 
$\overline{MS}$-scheme at the 3-loop order in Ref.[38]. In Ref.[27] 
the order $a_s^2$-term was related to the calculated 
in Ref.[26] light-by-light-type diagrams, contributing to the 3-point 
function of Eq.(50).

Let us now consider the conformal-invariant limit of all discussed 
in this Section results. In this case we have the Crewther-type 
relation 
\begin{equation}
C_{EJ}^{SI}(a_s(Q^2))C_D^{SI}(a_s(Q^2))|_{c-i}=1~~~.
\label{crew3}
\end{equation}
We do not still know, whether there exist its QCD 
generalization, analogous to the one of Eq.(28), discovered in 
Ref.[5]. From the general grounds we expect, that both sides of 
Eq.(58) should be multiplied by the anomalous dimension term 
$AD$, defined in Eq.(53). It is interesting to study the problem of 
the possibility of the existence of any relations between the extra 
light-by-light-type terms $\Delta C_D^{SI}(a_s)$, $\Delta 
C_{EJ}^{SI}(a_s)$ and the light-by-light-type QCD contributions to 
the r.h.s. of Eq.(58). The most surprizing fact is that the 
explicite analytical expression for $\Delta C_{EJ}^{SI}(a_s)$ 
contains the $\zeta(3)$-term~\cite{Larin2}, while the 3-loop 
correction $\Delta C_D^{SI}(a_s)$ does not contain it (see the 
first work of Ref.[49]). We think that the calculations 
of the NNLO corrections to the Ellis-Jaffe sum rule~\cite{planned} 
might be useful not only for  the 
comparison of their results with the scheme-invariant estimates of 
Ref.[88], but for  the analysis of the possibility of
the existence of new ``wonders'' in the non-conformal deviations from 
the Crewther-type relation of Eq.(58).  

\newpage

\section*{Acknowledgments}

We are grateful to the colleagues from Theoretical Division of
the INR where the author of this talk was continuously growing up
to the invitation to participate in the Workshop ``Continuous
Advances in  QCD''.

It is the pleasure to acknowledge the very constructive question of
R.L. Jaffe, which initiated our interest to the Crewther relation.

We wish to thank D.J. Broadhurst, S.J. Brodsky, G.T. Gabadadze and H.J. Lu
for their contribution to our productive collaborative research, aimed to 
the study of the different
aspects of the Crewther relation and its QCD generalizations and once more 
G.T. Gabadadze for his useful comments, related to the subjects, discussed 
in this work.

We express our warm gratitude to R. Jackiw for informing us recently about
the old work of Ref.[22], related to the analysis of the
consequencies of the conformal-invariant variant of this fundamental relation
and to all members of the Organizing Committee of this productive Workshop 
for the invitation and hospitality in Minneapolis.

This work was supported in part by the Russian Foundation of the Basic
Research, Grants N 96-02-18897 and N 96-01-01860 and is done within
the program of the INTAS project N 93-1180.

\newpage


\end{document}